 \documentclass[12pt]{article}
\usepackage{amsmath,amssymb}
  
\usepackage{bm}

\newcommand{\be}{\begin{equation}}

\newcommand{\ee}{\end{equation}}
\newcommand{\bea}{\begin{eqnarray}}
\newcommand{\eea}{\end{eqnarray}}

\newcommand{\nn}{\nonumber}

\newcommand{\bi}{\begin{enumerate}}
\newcommand{\ei}{\end{enumerate}}
\newcommand{\bref}[1]{(\ref{#1})}
 \newcommand{\B}{\beta} \newcommand{\gam}{\gamma}
 \newcommand{\D}{\delta} 
\newcommand{\ep}{\epsilon}

\newcommand{\lam}{\lambda}        \newcommand{\s}{\sigma}
          
\newcommand{\z}{\zeta}          
\newcommand{\h}{\eta}

\def\6{\partial}\def\7{\tilde}\def\8{\hat}

\def\pa{\partial}

\def\CL{{\cal L}}

\def\vs{\vskip 4mm}\def\={{\;=\;}}\def\+{{\;+\;}}
\def\NLR{{non-linear realization} }
\def\Euclidean{{euclidean}}
\def\Poincare{{Poincar\'e }}
\begin{document}

%\subheader
{\hfill {\rm ICCUB-10-174,UB-ECM-PF 10/42}}
\vskip 2cm %\title
\begin{center} 
{\huge SuperParticle realization of twisted ${\cal N}=2$ SUSY algebra}
\vskip 2cm 
{\Large
%\author[a]\author[b]\author[c]
{Roberto Casalbuoni${}^a$}, {Joaquim Gomis${}^b$} and {Kiyoshi Kamimura${}^c$}
%\affiliation[a]\affiliation[b]\affiliation[c]
}\vskip 1cm
{\it 
{${}^a$Department of Physics, University of Florence and INFN Via G. Sansone 1, 50019 Sesto Fiorentino (FI) and GGI, Florence, Italy}

{${}^b$Departament ECM and ICCUB, Universitat de Barcelona, Diagonal 647, 08028 Barcelona, Spain}

{${}^c$Department of Physics, Toho University, Funabashi, Chiba 274-8510, Japan}
\vskip 5mm

{\it E-mails: casalbuoni@fi.infn.it, gomis@ecm.ub.es, kamimura@ph.sci.toho-u.ac.jp} 
%\emailAdd{casalbuoni@fi.infn.it}\emailAdd{gomis@ecm.ub.es}\emailAdd{kamimura@ph.sci.toho-u.ac.jp} \keywords{\today}
}\end{center}
\vskip 3cm

\abstract
{We construct a pseudoclassical particle model associated to the twisted ${\cal N}=2$ SUSY algebra in four dimensions. The particle model has four kappa symmetries. Three of them
can be used to reduce the model to the vector supersymmetry particle case. The quantization of the model gives rise to two copies of the 4d Dirac equation. The kappa symmetries result to be associated to 4 TSUSY invariant bilinear odd operators who are null operators when a particular condition is satisfied. These operators are in correspondence one to one with analogous operators existing in the case of the ${\cal N}=2$ SUSY algebra, making both cases 1/2 BPS.}

%\pacs{12.38.-t, 26.60.+c, 74.20.-z, 74.20.Fg, 97.60.Gb}

%\maketitle
\eject

\section{Introduction \label{sec:0}}

Supersymmetry plays a crucial role in field theories, supergravities
and String/M theory. In flat space-time supersymmetry is
characterized by the presence of odd spinor charges that together
with the generators of the \Poincare group form the target space Super
\Poincare group.

There is also  an odd vector extension of the
\Poincare group, called the Vector Super \Poincare (VSUSY) \cite{Barducci:1976qu}.
Witten \cite{Witten:1988ze} introduced topological
$\mathcal{N}=2$ Yang-Mills theories by performing a topological twist in {\Euclidean} 4 dimensional space-time.
After this twist, the fermionic generators become a vector, a scalar and
an anti-selfdual tensor \cite{Alvarez:1994ii,Kato:2005fj}. After
truncation of the anti-selfdual sector, the twisted supersymmetry (TSUSY) algebra coincides with the {\Euclidean} VSUSY algebra.

A pseudoclassical particle model invariant under Super \Poincare was introduced in
\cite{Casalbuoni:1976tz} and extended later to  ten dimensions          \cite{Brink:1981nb}. In this last case the lagrangian
consists of
the Nambu-Goto
and the Wess-Zumino terms. 
Choosing a particular set of values of their coefficients, a fermionic gauge
symmetry called kappa symmetry \cite{deAzcarraga:1982dw}
\cite{Siegel:1983hh} appears. In the case of  $\mathcal{N}=2$ Super \Poincare in four dimensions (4d)
the number of independent kappa symmetries is four and therefore the superparticle is 1/2 BPS.
The covariant quantization of this model has several difficulties.

A pseudoclassical particle  invariant under VSUSY \Poincare is the spinning particle
introduced in \cite{Barducci:1976qu}. The quantization preserving VSUSY gives rise to  two copies of
the 4d Dirac equation \cite{Casalbuoni:2008iy}. On the other hand, breaking the rigid
supersymmetry by a suitable constraint on the Grassmann variables, one
 recovers the 4d Dirac equation of
\cite{Barducci:1976qu,Brink:1976uf,Brink:1976sz,
Berezin:1976eg}.
The lagrangian  contains a Dirac-Nambu-Goto piece and two
Wess-Zumino terms.
 For particular values of the coefficients of the lagrangian,
the model has world line gauge supersymmetry which is analogous to
the fermionic kappa symmetry of the superparticle case. When we
require world-line supersymmetry the model has bosonic BPS
configurations that preserve 1/5 of the vector supersymmetry.
The fact that the model is 1/5 BPS makes it difficult to find a relation with the superparticle, since  this is  1/2 BPS.

In this paper we construct a pseudoclassical particle model in a 4d {\Euclidean} space which is invariant under the twisted $\mathcal{N}=2$ SUSY algebra. The variables of lagrangian are the space-time coordinates $x^\mu$ and the odd real Grassmann variables $\xi^\mu, \xi^5,\xi^{\mu\nu}$
that are a vector, a scalar and a self-dual tensor under 4d rotations.
The lagrangian  contains a Dirac-Nambu-Goto piece and two
Wess-Zumino terms.
For particular values of the coefficients of the lagrangian,
the model has {\it four} fermionic kappa symmetries. Three of them
allow
to eliminate the selfdual Grassmann variables $\xi^{\mu\nu}$ and the result is equivalent to the {\Euclidean} spinning particle model invariant under VSUSY. The quantization of the model
in the reduced space leads to two copies of the 4d {\Euclidean} Dirac equation.

The model has bosonic BPS
configurations that preserve 1/2 of the twisted $\mathcal{N}=2$ supersymmetry.
Note that the TSUSY model is 1/2 BPS like the superparticle.
This is expected on the basis of the correspondence of the TSUSY and $\mathcal{N}=2$ SUSY algebras and we will show explicitly this relation.

\section{TSUSY algebra \label{sec:1}}

We will consider the SUSY algebra for ${\cal N}=2$ in the {\Euclidean} case. The algebra of the odd generators is given by 
\be \{Q^i_\alpha,\tilde Q_{jA}\}=2(\sigma^\mu)_{\alpha A} P_\mu \delta^i_j, \qquad \{Q_\alpha^i,Q^j_\beta\}=\epsilon_{\alpha\beta}\epsilon^{ij} Z_1,\qquad \{\tilde Q_{iA},\tilde Q_{jB}\}=\epsilon_{AB}\epsilon_{ij}Z_2\ee
where $\s^\mu=(i\tau_j,1)$ and $\epsilon=i\tau_2$  and  {$\tau_i$ are the Pauli matrices}. The indices $i,\alpha,A$ describe the following groups
\be
i\in SU(2)^{\cal R},~~~\alpha\in SU(2)_L,~~~A\in SU(2)_R,\ee
where $SU(2)^{\cal R}$ is the $R$-symmetry group and $SU(2)_L\times SU(2)_R$ describes the
space-time rotation group in four dimensions, $O(4)$. This algebra can be twisted in the following way \cite{Witten:1988ze}, see also \cite{Alvarez:1994ii,Kato:2005fj}. We identify a new rotation group $O(4)_N$ with the product of the diagonal part of the product of $SU(2)^{\cal R}\times SU(2)_L$ times $SU(2)_R$. 
That is
\be
O(4)_N\approx [SU(2)^{\cal R}\times SU(2)_L]_{D}\times SU(2)_R=SU(2)_D\times SU(2)_R.\ee Where we have denoted by $SU(2)_D$ the diagonal subgroup of the direct product in parenthesis in the previous equation. In other words we identify the spinor indices of the internal symmetry $SU(2)^{\cal R}$ with the spinor indices of the space-time symmetry $SU(2)_L$.
Then, the indices $i,\alpha$ transform both under $SU(2)_D$, whereas $A$ under $SU(2)_R$. We have
\be
Q_\alpha^i\in (1/2,0)\times(1/2,0)=(0,0)\oplus (1,0),\quad
\tilde Q_{iA}\in (1/2,0)\times(0,1/2)=(1/2,1/2).\ee
It is important to notice that the two inequivalent spinorial representations in $O(4)$ (from now on $O(4)_N\to O(4)$) are both pseudoreal. We see that $Q_\alpha^i$ describes a scalar and a self-dual tensor, that is a spin 1, whereas $\tilde Q_{iA}$ a fourvector. This can be made explicit by decomposing the supercharges $Q$ and $\tilde Q$ 
as follows\footnote{For simplicity we suppress the indices whenever it does not create ambiguities.}
\be
Q= iG_5 -\frac {i}2\sigma^{\mu\nu}G_{\mu\nu}, ~~~~\tilde Q= i\,\sigma^\mu G_\mu.\label{eq:2.7}\ee
\vs

In terms of $(G_\mu, G_5, G_{\mu\nu})$ the twisted algebra
is given by
\be
\{G_5,G_5\}=\7Z,~~~\{G_5,G_\mu\}= -P_\mu,~~~\{G_\mu,G_\nu\}=Z\delta_{\mu\nu}\ee
\be
\{G_{\mu\nu},G_5\}=0,~~~\{G_{\mu\nu},G_{\rho\lambda}\}=4P^+_{\mu\nu,\rho\lambda}\7Z,~~~
\{G_{\mu\nu},G_\rho\}=-4P_{\mu\nu,\rho\lambda}^+P^\lambda\ee
\be
\left[G_{\mu},M_{\rho\s}\right]=-i\h_{\mu[\rho}G_{\s]},~~~
\left[G_{\mu\nu},M_{\rho\s}\right]=-i\h_{\nu[\rho}G_{\mu\s]}+i\h_{\mu[\rho}G_{\nu\s]},~~~\left[G_{5},M_{\rho\s}\right]=0.\ee
Here
\be \tilde Z=-\frac{Z_1}2,~~~~Z=-\frac{Z_2}2,\ee
and $P_{\mu\nu,\rho\s}^\pm$  project out the
 (anti)self-dual part of a tensor:
\be
P^{\pm}_{\mu\nu,\rho\lambda}=\frac 14(\delta_{\mu\rho}\delta_{\nu\lambda}-\delta_{\nu\rho}\delta_{\mu\lambda}\pm\epsilon_{\mu\nu\rho\lambda}).\ee
Notice that due to the self-dual properties of $\sigma_{\mu\nu}$, $G_{\mu\nu}$  is self-dual   with three independent components satisfying
 $G_{\mu\nu}=P^{+}_{\mu\nu,\rho\lambda}G^{\rho\lambda}$.

In Minkowski space SUSY algebra admits a $U(1)$ automorphism group generated by the ghost number. In the {\Euclidean} case such an automorphism does not exist due to the pseudo-reality of the spinors. On the other hand there is an automorphism given by the following scale transformation (we will call this ghost dilation to be distinguished from the automorphism generated by the usual dimensional scale  dilations). This is given by
\bea
P_\mu\to P_\mu,\quad M_{\mu\nu}\to M_{\mu\nu},&&
Z_1\to e^{+2\lambda}Z_1,\quad Z_2\to e^{-2\lambda}Z_2\nn\\
Q_\alpha^i\to e^{+\lambda} Q_\alpha^i,&&\tilde Q_{iA}\to e^{-\lambda}\tilde Q_{iA}.
\eea
Clearly this symmetry of the algebra is related to the ghost symmetry through the analytic continuation $\lambda\to i\lambda$. The corresponding transformation properties  of the scalar, vector and tensor generators are
\be
G_5\to e^{+\lambda}G_5,~~
G_\mu\to e^{-\lambda}G_\mu,~~G_{\mu\nu}\to e^{+\lambda}G_{\mu\nu}.\ee
Notice that, in Minkowski space the BPS operators are constructed (in the rest frame) by taking linear combinations of $Q$ and $ Q^\dagger$. However, due to the pseudoreality condition, in the euclidean case one has to take linear combinations of $Q$ and $\tilde Q$. The total number of BPS operators in TSUSY is expected to  be 4 as in ${\cal N}=2$ SUSY.  However, since the BPS charges should be expressed in terms of the  generators $G_5,G_\mu$ and $G_{\mu\nu}$, it is easier to work in an arbitrary Lorentz frame. We should be  able to construct 4 linear combinations, one involving $G_5$ and $G_\mu$ and the other $G_\mu$ and $G_{\mu\nu}$. This is because $G_5$ and $G_{\mu\nu}$ depend on $Q$ whereas $G_\mu$ depends on $\tilde Q$. We then  construct the following four combinations for $P^2\not=0$
\be
B=P^\mu G_\mu +Z G_5,~~~B_\mu=G_{\mu\nu}P^\nu+\tilde Z P^\bot_{\mu\nu}G^\nu, \label{eq:5.1}\ee
where
\be P^\bot_{\mu\nu}=\delta_{\mu\nu}-\frac{P_\mu P_\nu}{P^2}.\ee
These operators are Lorentz covariant and   scale correctly under  the dilation automorphism discussed previously. Notice that  the transverse projector ensures that only three operators out of $B_\mu$ are independent.
Let us now evaluate the anticommutators of these operators. We get
\be
\{B,B\}=-(P^2-Z\tilde Z)Z,~~~~~~\{B_\mu,B_\nu\}=-(P^2-Z\tilde Z)\tilde Z P_{\mu\nu}^\bot,~~~~~\{B,B_\mu\}=0.\label{eq:16}\ee
Let us also  evaluate the anticommutators with the odd generators of TSUSY. We find
\be
\{G_\mu,B\}=\{G_{\mu\nu},B\}=0,~~~~\{G_5,B\}=-(P^2-Z\tilde Z),\ee
\be
\{G_\mu,B_\nu\}=-(P^2-Z\tilde Z)P_{\mu\nu}^\bot,~~~\{G_5, B_\mu\}=\{G_{\mu\nu},B_\rho\}=0.\ee
Therefore, when the condition
\be P^2=Z\tilde Z\label{eq:BPS}\ee
is satisfied, the operators $B$ and  the three independent components of $B_\mu$ are four null operators  invariant under TSUSY transformations. The states annihilated  by the operators $B$, $B_\mu$ are the BPS states.
In the rest frame one gets\be
B=-\frac {i}2\left(m{\rm Tr}(\tilde Q)+Z{\rm Tr}(Q)\right),\ee
\be
B_4=0,~~~B_i=\frac12\left(m{\rm Tr}(\tau_i Q)-\tilde Z{\rm Tr}(\tau_i\tilde Q)\right).\ee
From these expressions one recovers easily the usual expressions for the BPS operators in the language of SUSY except for the factors $m$ and $Z$, $\tilde Z$.
In fact, in Minkowski space, in order to get BPS operators with definite ghost number  it is enough to take the hermitian conjugate of $Q$. Since the $U(1)$ generated by the ghost number goes into dilation group, one needs to insert appropriate factors $Z$, $\tilde Z$, since $Q$ and $\tilde Q$ scale in a different way. However, $Z$ and $\tilde Z$ bring dimensions requiring the insertion of appropriate powers of mass.

When the  condition (\ref{eq:BPS}) is satisfied we can evaluate both $G_5$ and $G_{\mu\nu}$ using the equations $B=B_\mu=0$. We get
\be G_5=-\frac 1 ZP_\mu G^\mu, ~~~~G_{\mu\nu}=\frac 4 ZP^+_{\mu\nu,\rho\lambda}P^\rho G^\lambda.\label{eq:21}\ee
In the rest frame we have
\be G_5=-\frac m Z G_4,~~~~ G_{i4}=-\frac m Z G_i\ee with the further duality condition
\be
G_{ij}=\epsilon_{ijk}G_{k4}.\ee

In the next section we will study the issue of realizing the TSUSY algebra on the phase space of a single particle. In principle, the particle model is described by the position variables and by 8 real Grassmann variables associated to the odd generators of TSUSY, $\xi_5$, $\xi_\mu$ and $\xi_{\mu\nu}$, where the last variables form a self-dual tensor.
If one chooses the parameters of the action in such a way that  the realization of the  condition (\ref{eq:BPS}) is satisfied, one expects that the number of the Grassmann variables describing the  model can be reduced.
This is because  $B$ and $B_\mu$ become null operators. In the particle model the corresponding expressions will vanish giving rise to constraints. On the other hand, the equations (\ref{eq:16}) show that these operators anticommute with each other. As a consequence, at the level of the particle model they give rise to 4 first-class constraints. As it is well known,
first class constraints generate gauge transformations (kappa-symmetries in the case of SUSY) and this allows to eliminate one  odd variable in configuration space for each transformation via gauge fixing. For example,  one can fix three gauge conditions and eliminate the variables $\xi_{\mu\nu}$. In this case one obtains the VSUSY particle model of reference \cite{Casalbuoni:2008iy}. As a further step it is possible to use the remaining constraint to eliminate $\xi_5$.  The possibility of eliminating the variables $\xi_5$ and $\xi_{\mu\nu}$ goes together with the fact that, when the condition (\ref{eq:BPS}) is satisfied, the generators $G_5$ and $G_{\mu\nu}$ can be expressed through $G_\mu$ (see eq. (\ref{eq:21})). In the euclidean case, this means that the model involving only $\xi_\mu$ has the full TSUSY invariance. However, in  Minkowski space this does not happen because the self-dual generators $G_{\mu\nu}$, and the corresponding parameters $\xi_{\mu\nu}$, lose the reality property.

The mechanism discussed here looks completely general, that is, when the choice of a particular realization of the algebra gives rise to null operators, in the associated particle model there are constraints. The character of these constraints (first or second class) will depend on the algebraic relations existing among the null operators. In the present case, since  the 4 null operators anticommute among themselves, they give rise to 4 first-class constraints.

\section{TSUSY particle \label{sec:2}}

To construct a pseudoclassical  model invariant under TSUSY we will use the method of \NLR \cite{Coleman:1969sm}.
The starting point is to consider the Maurer-Cartan (MC) form associated to the twisted {\Euclidean} superspace,  $\frac{G}{H}=\frac{{\rm TSUSY}}{SO(4)}$ which locally is parametrized by
\be g_0=e^{iP_\mu x^\mu} e^{\frac{i}4G_{\mu\nu}\xi^{\mu\nu}} e^{iG_5\xi^5}e^{iG_\mu \xi^\mu}e^{iZc} e^{i\7Z \7c}.\label{eq:cosetTS0}\ee
The MC 1-form of the twisted superspace
is \bea \Omega_0&=&-i{g_0}^{-1}dg_0 =P_\mu \tilde L_P^\mu+G_5 \tilde L^5_G +G_\mu{\tilde L_G^\mu}+\frac14G_{\mu\nu}{\tilde L_G^{\mu\nu}}+Z\tilde L_Z+\tilde Z \tilde L_{\7Z}. \eea The even components of the  1-form are given by \be
\tilde L_P^\mu=dx^\mu- i \xi^\mu d\xi^5+i\xi_\s d\xi^{\mu\s}
\nn\ee \be \tilde L_Z=dc+\frac i 2 \xi^\mu d\xi_\mu,~~~\tilde L_{\tilde Z}=d\tilde c+\frac i 2\xi^5 d\xi^5+\frac i 8\xi_{\mu\nu}d\xi^{\mu\nu}
\label{1formb}\ee whereas the odd ones are \be
\tilde L_G^\mu=d\xi^\mu,\qquad \tilde L_G^5=d\xi^5,\qquad
\tilde L_G^{\mu\nu}= d\xi^{\mu\nu}. \label{1formf0}\ee

The TSUSY transformations of the  real superspace coordinates are given by
\bea
G_{\mu\nu}&:& \D\xi^{\mu\nu}=\B^{\mu\nu},\quad
\D\7c=\frac{i}8\xi_{\mu\nu}\B^{\mu\nu},
\nn\\
G_{5}&:& \D\xi^{5}=\B^{5},\quad \D\7c=\frac{i}2\xi^5\B^5,
\nn\\
G_{\mu}&:& \D\xi^{\mu}=\B^{\mu},\quad \D x^\mu=-i\xi^5\B^\mu+i\xi^{\mu\nu}\B_\nu, \quad \D c=\frac{i}2\xi_\mu\B^\mu,
\nn\\
P_{\mu}&:& \D x^{\mu}=\ep^{\mu},
\nn\\
Z&:& \D c=\ep_Z,
\nn\\
\7Z&:& \D \7c=\ep_{\7Z}.
\label{tsusytrans}\eea
These transformations leave invariant the above MC 1-forms.

Now we want to consider the motion of a massive particle in this space. Then, the natural coset is $\frac{G}{H}=\frac{{\rm TSUSY}}{SO(3)}$, regarding $x^4$ as "euclidean time ".
Since the "{\Euclidean}  boosts", $M_{4i}$
 are broken  spontaneously by the presence of the particle,  see for example \cite{Gauntlett:1989qe,Gomis:2006xw} for the case of relativistic particles with Minkowski signature. Then the elements of the coset are of the form
\be g=g_{0}\,U,\qquad \quad U\equiv\;e^{iM_{4i}v^i},\label{Beq:coset} \ee
where $v^{i}$ are the parameters associated to the "{\Euclidean} boost".
The corresponding MC form is
\bea
\Omega&=&-ig^{-1}dg=
U^{-1}\Omega_0\,U+U^{-1}dU \nn\\
&=&P_\mu L_P^\mu+\frac12M_{\mu\nu} {L_M^{\mu\nu}}+G_5 L^5_G +G_\mu{L_G^\mu}+\frac14G_{\mu\nu}{L_G^{\mu\nu}}+ZL_Z+\tilde Z L_{\7Z},
\eea
with
\be
L_P^\mu={\Lambda^\mu}_\nu \tilde L_P^\nu,\qquad L_M^{\mu\nu}= {\Lambda^\mu}_\rho d{\Lambda^\nu}_\s\h^{\rho\s},\qquad
L_Z=\tilde L_Z,\qquad L_{\tilde Z}=\tilde L_{\tilde Z}\label{1formb2}, \ee  \be
L_G^\mu={\Lambda^\mu}_\nu \tilde L_G^\nu,\qquad L_G^5=\tilde L_G^5,\qquad
L_G^{\mu\nu}={\Lambda^\mu}_\rho {\Lambda^\nu}_\s \tilde L_G^{\rho\s}. \label{1formf}\ee
where we have used
\bea
U^{-1}\;P_\nu\; U&=& P_\mu{\Lambda^\mu}_\nu(v),\qquad U^{-1}\;M_{\mu\nu}\; U= M_{\rho\s}{\Lambda^\rho}_\mu(v){\Lambda^\s}_\nu(v).
\eea 
${\Lambda^\mu}_\nu(v)$ is a finite 4d rotation matrix depending on  the {\Euclidean} boost
parameters $v^{i}$.

There are three even 1-forms invariant under the left transformations, $
L_P^4,\, L_Z,\, L_{\7Z}.$ Then, the simplest form of the particle lagrangian  is a linear combination of these invariant one forms:
\bea
\CL\,d\tau&=&-[{m}\,L^4_P+\B \,L_Z+\gamma \, L_{\7Z}]\,\tilde{} \nn\\
&=&-\left({m}\,{\Lambda^4}_\nu (\dot x^\nu- i \xi^\nu \dot \xi^5+ i \xi_\s \dot \xi^{\nu\s})+\B \,(\dot c+\frac i 2 \xi^\mu \dot \xi_\mu)+\gamma \, (\dot{\tilde c}+\frac i 2\xi^5 \dot \xi^5+\frac i 8\xi_{\mu\nu}\dot \xi^{\mu\nu})\right)d\tau,\nn\\ \label{lagabcTS}\eea
where the $[\cdots]\,\tilde{}$ denotes  the pullback to the world line forms
: $[dx^\mu]\,\tilde{} \equiv \dot x^\mu(\tau)\, d\tau$, etc.
If we introduce the momenta $p_\mu=-{m}\,{\Lambda^4}_\mu$ we can write the canonical lagrangian  as
\be
\CL_1=p_\nu (\dot x^\nu- i \xi^\nu \dot \xi^5+ i \xi_\s \dot \xi^{\nu\s})-\frac{e}{2}(p^2-{m}^2)-\B \,\frac i 2 \xi^\mu \dot \xi_\mu-\gamma( \,\frac i 2\xi^5 \dot \xi^5 +\frac i 8\xi_{\mu\nu}\dot \xi^{\mu\nu}),
\label{lagabcpTS}\ee
where we have removed total derivative terms.

\section{Constraints and Kappa Symmetries of TSUSY Lagrangian}

In this section we will compute the constraints and the generators of the gauge symmetries for the TSUSY particle.
The momenta for the Lagrangian \bref{lagabcTS}
are given by\footnote{the momenta of the fermionic variables are computed using right derivative to define odd momenta then $H=\zeta\dot\xi-L$ and
$\{\xi,\zeta\}=+1$.}
\bea
p_\mu&=&-{m}\,{\Lambda^4}_\mu,\qquad p_{v^i}=0, \label{momxmu1}\\
p_c&=&-\B,\qquad p_{\7c}=-\gamma, \label{momxmu2}\\
\z_\mu&=&-\B\frac{i}2\xi_\mu,\qquad \qquad \qquad \qquad  \label{momxmu4}\\
\z_5&=&-ip_\mu\xi^\mu-\gamma\frac{i}2\xi^5, \label{momxmu3}\\
\z_{\mu\nu}&=&4\,i\,p_\rho\xi_\s {P^{\rho\s}}_{,\mu\nu}-\gamma\frac{i}2\xi_{\mu\nu},\qquad
\label{momxmu5} \eea
where the basic Poisson brackets of the fermionic variables are
\be
\{\xi^{\mu},\z_{\nu}\}= {\D^{\mu}}_{\nu},\quad \{\xi^{\mu\nu},\z_{\rho\s}\}= 4\,{P^{\mu\nu}}_{,\rho\s},\quad \{\xi^{5},\z_{5}\}=1.
\ee
We have seven even constraints  \bref{momxmu1} , one first class
\be
\phi=\frac12(p^2-\,m^2),\label{masscon}\qquad
\ee
and six second class that can be used to  eliminate $(v^i,p_{v^i}),$
\be
p_{v^i}=0,\qquad v^i=v^i(p_\mu).
\ee
There are also two even constraints  \bref{momxmu2} that are first class generating local shift of $c$ and $\7c$,
\be
\chi_c=p_c+\B=0,\qquad \chi_{\7c}=p_{\7c}+\gamma=0. 
\ee
We have four fermionic constraints \bref{momxmu4} which are second class
\be \chi_\mu=\z_\mu+\B\frac{i}2\xi_\mu=0. \label{conmom4}\ee
 For second class constraints $\phi_\alpha=0$ we can either define the Dirac bracket
\be \{A,B\}^*=\{A,B\}-\{A,\phi_\alpha\}c^{-1}_{\alpha\beta}\{\phi_\beta,B\}
\ee or equivalently introduce the $A^*$ variables associated to any $A$, as, see for example
\cite{Henneaux:1992ig}
\be
A^*=A-\{A,\phi_\alpha\}c^{-1}_{\alpha\beta}\phi_\beta
\ee
where $\{\phi_{\alpha},\phi_\beta\}=c_{\alpha\beta}$.
 The relations 
\bref{momxmu3} and   \bref{momxmu5} give the odd constraints
\bea
\chi_5&=&\z_5+ip_\mu\xi^\mu+\gamma\frac{i}2\xi^5=0, \label{conmom3}\\
\chi_{\mu\nu}&=&\z_{\mu\nu}-4\,i\,p_\rho\xi_\s {P^{\rho\s}}_{\mu\nu}+
\gamma\frac{i}2\xi_{\mu\nu}=0
\eea
for which $\chi_5^*, \chi_{\mu\nu}^*$ are\bea
\chi_5^*&=&
\z_5+\gamma\frac{i}2\xi^5-\frac1{\B}p_\mu(\z_\mu-\B\frac{i}2\xi_\mu)=0, \label{conmom3s}\\
\chi_{\mu\nu}^*&=&
\z_{\mu\nu}+\gamma\frac{i}2\xi_{\mu\nu}+\frac{4}{\B}\,p_\rho(\z_\s-\B\frac{i}2\xi_\s){P^{\rho\s}}_{\mu\nu}=0.
\eea
They are  first class constraints when the condition
\be m^2=\beta\gamma\label{kappacon}\ee
is satisfied.
In fact,  they satisfy the following Poisson brackets
\bea
\{\chi^*_5,\chi^*_5\}&=&-\frac{i}\B\;(p^2-\B\gamma)=
-\frac{i}\B\;(2\phi+m^2-\B\gamma),\\
\{\chi^*_5,\chi^*_{\mu\nu}\}&=&\frac{4i}\B\;p_\lam p_\rho {P^{\rho\lam}}_{,\mu\nu}=0,\nn\\
\{\chi^*_{\mu\nu},\chi^*_{\rho\s}\}&=&
%\gamma {i}\,4P_{\mu\nu\rho\s}-16 p^\A P_{\A\gam\mu\nu} p^\B P_{\B\D\rho\s}\frac{i}\B \h^{\gam\D} %\nn\\&=&-4\frac{i}\B\;(p^2-\B\gamma)P_{\mu\nu\rho\s}=
-4\frac{i}\B\;(2\phi+m^2-\B\gamma)P_{\mu\nu,\rho\s}.
\eea
Notice that $\chi_{\mu\nu}$ is a self-dual tensor, $P^{-}_{\mu\nu,\rho\lambda}\chi^{\rho\lambda}=0$.
In total the number of independent kappa symmetries is four like in the case of the ${\cal N}=2$ superparticle,  as expected from the discussion made at the end of Section \ref{sec:1}.

The kappa transformations are generated by
\be
G=\chi_5^*\kappa^5+\frac14\chi_{\mu\nu}^*\kappa^{\mu\nu},
\ee
from which
\bea
 \D\xi^5&=&\kappa^5,\quad \D\xi^{\mu\nu}=\kappa^{\mu\nu},\quad 
\D\xi^{\mu}=\frac{1}{\B}({p}^\mu\,\kappa^5+{p}_\nu\kappa^{\nu\mu}),\quad 
\D x^\mu= %i\xi^\mu\kappa^5-\,i {P^{\mu}}_{\nu\rho\s}\,\xi^\nu\kappa^{\rho\s}=
i\xi^\mu\kappa^5+\,i \xi_\nu\kappa^{\nu\mu}.
\eea The lagrangian \bref{lagabcTS} transforms as a total divergence
when \bref{kappacon} holds.

The canonical generators of the  TSUSY transformations
(\ref{tsusytrans}) are
\bea
P_{\mu}&=& p_{\mu},\qquad Z= p_c,\qquad  \7Z=p_{\7c}, \qquad
G_{\mu\nu}=\z_{\mu\nu}+p_{\7c}\frac{i}2\xi_{\mu\nu},
\nn\\
G_{5}&=& \z_{5}+p_{\7c}\frac{i}2\xi^5,
\qquad
G_{\mu}= \z_{\mu}-i\xi^5p_\mu+ip^\nu\xi_{\nu\mu}+p_c\frac{i}2\xi_\mu.
\eea
In terms of the canonical variables the quantities  $B$ and $B_\mu$ of equation (\ref{eq:5.1}) are given by
\bea
B&=&P^\mu G_\mu +Z G_5=
\nn\\&=&
\chi_c\,\frac{i}2p^\mu\xi_\mu +\chi_c\,(\z_{5}+p_{\7c}\frac{i}2\xi^5)-\B\,\,\chi_5^*
-i\xi^5(p^2-\B\gamma),\\
B_\mu&=&G_{\mu\nu}P^\nu+\tilde Z P^\bot_{\mu\nu}G^\nu=
\nn\\&=&
\chi_{\7c}\,\frac{i}2\xi_{\mu\nu}p^\nu+\chi_{\7c}\,
P^{\bot\nu}_{\mu}(\z_{\nu}+ip^\rho\xi_{\rho\nu}+(\chi_{c}-\B)\frac{i}2\xi_\nu)
-\chi_{c}\gamma P^{\bot\nu}_{\mu}\frac{i}2\xi_\nu+
\nn\\&&+\chi_{\mu\nu}^*p^\nu+\frac{1}{\B}\,P^{\bot\nu}_{\mu}
(\z_\nu-\B\frac{i}2\xi_\nu)(p^2-\B\gamma).
\label{eq:5.11}\eea
Then, if we use the  condition $P^2-Z\7Z=p^2-\B\gam=0$ the $B,B_\mu$
are linear combinations of the first class constraints. Their expressions, barring the trivial constraints
$\chi_c=\chi_{\7c}=0$,  are
\be
B=-\B\,\,\chi_5^*,\qquad
B_\mu=\chi_{\mu\nu}^*p^\nu.
\ee

\section{Gauge fixed lagrangian and quantization on the reduced space}

Taking into account the first class constraints we have obtained we can introduce the
gauge fixing conditions \be \xi^5=\xi^{\mu\nu}=c=\7c=0.\ee
The gauge fixed Lagrangian of \bref{lagabcTS} is
\bea
\CL^{gf}&=&-{m}\,{\Lambda^4}_\mu \dot x^\mu-\B \,\frac i 2 \xi^\mu \dot \xi_\mu
=p_\mu \dot x^\mu-\frac{e}{2}(p^2-\,{m}^2)-\B \,\frac i 2 \xi^\mu \dot \xi_\mu.
\label{lagabcTSgf}\eea
 The global symmetry generators are, including compensating gauge transformations,
\bea
G_\mu^*&=& \z_\mu-\B\frac{i}2\xi_\mu,\qquad
G_5^*=\frac1\B p^\mu(\z_\mu-\B\frac{i}2\xi_\mu),\nn\\
G_{\mu\nu}^*&=&-\frac{4}{\B}\,p_\rho(\z_\s-\B\frac{i}2\xi_\s){P^{\rho\s}}_{,\mu\nu}.\label{Gmunus}\eea

In the Lagrangians \bref{lagabcTSgf} the self-dual real coordinates $ \xi^{\mu\nu}$ have disappeared thoroughly. Now we can use the lagrangian \bref{lagabcTSgf} in Minkowski space,
by replacing $m^2\to -m^2$.
However, the global $G_{\mu\nu}^*$ transformations \bref{Gmunus}, realized in the Euclidian metric, 
are not symmetries in  Minkowski space due to the complex nature of the self-dual tensor transformation parameters $\ep^{\mu\nu}$.

Going to the Minkowski space and  fixing the gauge in relation to the reparametrization invariance
\be
x^0=\tau.
\ee
we can solve the constraint \bref{masscon} for $p_0$
\be
p_0=\pm \sqrt{{\vec p}^{\,2}+m^2}.
\ee
 The canonical form of the lagrangian \bref{lagabcTSgf} in
the reduced space\footnote{ An analogous discussion for the
lagrangian of spinning particle in \cite{Barducci:1976qu}
 was done in
reference \cite{Casalbuoni:2008iy}.} becomes \be L^{C*}= \pm
\sqrt{{\vec p}^{\,2}+m^2} +\,\vec{p}\,\dot{\vec{x}}
-\beta\frac{i}2\xi_\mu\dot\xi^\mu.\label{lagcf} \ee

Now we quantize the model. The basic canonical (anti-)commutators
are \be [x^i, p_j]=i {\delta^i}_j,\qquad
[\xi^\mu,\xi^\nu]_+=-\frac{1}\beta{\eta^{\mu\nu}}.\label{qcomred}\ee
Notice that the ambiguity in the sign the energy must be taken  into
account in the quantum theory. The hamiltonian for the lagrangian
\bref{lagcf} is the operator \be P_0=
\begin{pmatrix}{\omega}&0\cr 0&-{\omega}\end{pmatrix},
\qquad {\omega}\equiv\sqrt{{\vec p}^{\,2}+m^2}
,\label{P0}\ee
with eigenvalues $\pm {\omega}$.
The Schr\"odinger equation becomes
\be i\pa_\tau \Psi(\vec x,\tau)=P_0 \Psi(\vec x,\tau),\qquad
\Psi=\begin{pmatrix}\Psi_+\cr\Psi_- \end{pmatrix},
\label{Schgaugefixed}\ee
where $\Psi_+$ and $\Psi_-$ are positive and negative energy states.
 The odd real variables $\xi^\mu$ in \bref{lagcf}   must commute with all the bosonic variables, in
particular with the energy $P_0$ in \bref{P0}. Therefore they must be realized in terms of 8-dimensional gamma matrices \be
\xi^\mu= \sqrt{\frac{-1}{2\beta}}\;
\begin{pmatrix}\gamma^\mu&0\cr0&-\gamma^\mu \end{pmatrix}
\begin{pmatrix}\gamma^5&0\cr0&-\gamma^5 \end{pmatrix}
,\label{eq:5.8} \ee where the $\gamma^\mu$ and $\gamma^5$ are the
ordinary 4-component gamma matrices in 4-dimensions and
$\Psi$ is an 8 component wave function.
In \cite{Casalbuoni:2008iy} we have shown that the equations of motion are equivalent to two set of 4-component Dirac equations using an inverse Foldy-Whouthuysen
transformation.

\section{BPS Configurations}

Here we will consider the BPS equations for the TSUSY
particle. The corresponding  bosonic supersymmetric configurations
appear only when the lagrangians have the four kappa symmetries, i.e., when
$m^2=\B\gamma$.
Now we look for supersymmetric bosonic configurations. For
consistency we look for transformations of the fermionic variables
not changing their initial value that is supposed to vanish
\bea
0=\left.\delta\xi^5\right|_{\rm fermions=0}&=& \epsilon^5+
\kappa^5,\\
 0=\left.\delta\xi^\mu\right|_{\rm fermions=0}&=&\epsilon^\mu+ \frac{1}{\B}({p}^\mu\,\kappa^5+{p}_\nu\kappa^{\nu\mu}),\label{eq:71}
\\
0=\left.\delta\xi^{\mu\nu}\right|_{\rm fermions=0}&=&\epsilon^{\mu\nu}+\kappa^{\mu\nu}.
\eea
Notice that all $\epsilon^\mu$ can be expressed in terms of $\epsilon^5$ and the three
independent components of $\epsilon^{\mu\nu}$.
The consistency of equation (\ref{eq:71})  implies
\be
p^\mu= {\rm constant}
\ee
which is a 1/2 BPS configuration. Notice that the BPS solutions satisfy the Euler-Lagrange equations of motion.

\section{Discussions \label{sec:5}}

In this paper we have considered the twisted version of ${\cal N}=2$ SUSY. The corresponding algebra (TSUSY) contains odd tensorial generators, a scalar, a fourvector and a self-dual tensor. We have shown that there are four odd quadratic expressions in the generators , invariant under odd supertranslations, being null operators when the condition (\ref{eq:BPS}) is satisfied. These operators are one scalar, $B$, and one fourvector, $B_\mu$, with only three independent components (since it is orthogonal to the four-momentum). Requiring the vanishing of the last three independent operators, the TSUSY algebra reduces, as the independent generators are concerned, to the algebra of VSUSY (see eq. \cite{Casalbuoni:2008iy}).

By using the usual methods of the non linear realizations we have constructed the particle model associated to the TSUSY algebra (still remaining in an euclidean space). The natural variables describing the model are the parameters of the coset
space $\frac{{\rm TSUSY}}{SO(4)}$, that is the position $x_\mu$ and the odd  real quantities
$\xi_5$, $\xi_\mu$ and the self-dual tensor $\xi_{\mu\nu}$. The most simple invariant lagrangian turns out to depend on 3 parameters. The condition leading to four null quadratic operators in the generators of the algebra is equivalent to a condition among these three parameters. When this condition is satisfied, the model
 acquires four first-class constraints (corresponding to the vanishing of the operators $B$ and $B_\mu$). As a consequence there are four local symmetries, "kappa"-symmetries, allowing to eliminate the variables $\xi_5$ and $\xi_{\mu\nu}$.
This process is just in correspondence with the analogous procedure that in the case of ${\cal N}=2$ SUSY, allows the elimination of four fermionic variables. After this reduction, the euclidean model still has the full TSUSY invariance (with the generators $G_5$ and $G_{\mu\nu}$ expressed non-linearly in terms of $G_\mu$). At this stage we can also continue the model to the Minkowski space, however the complete TSUSY invariance is lost, due to the complex nature of the self-dual transformation parameters $\epsilon_{\mu\nu}$.

\vskip 3mm
{\bf Acknowledgments}

 We acknowledge discussions with Tomas Ort\'{\i}n, Alfonso Ramallo, Laura Tamassia and Toine Van Proeyen. We acknowledge partial financial support from projects FP2007-66665, 2009SGR502, CPAN( Consolider CSD2007-00042).

\end{document}